# Spin dynamics in the pyrochlore iridate, $Er_2Ir_2O_7$, investigated by µSR spectroscopy

R.H. Colman,[1,*] K. Vlášková,[1] A. Berlie,[2] and M. Klicpera[1]

[1] *Charles University, Faculty of Mathematics and Physics, Department of Condensed Matter Physics, Ke Karlovu 5, 121 16 Prague 2, Czech Republic*

[2] *ISIS Neutron and Muon Source, Science and Technology Facilities Council, Rutherford Appleton Laboratory, Didcot, OX11 0QX, United Kingdom*

*Corresponding author: ross.colman@mag.mff.cuni.cz*

**Abstract:**

Building a full understanding of the complex magnetism in the rare-earth pyrochlore iridates, $A_2Ir_2O_7$, is an ongoing issue in condensed matter physics. The possibility of two interpenetrating and interacting frustrated magnetic pyrochlore sublattices, connected with the strong spin-orbit coupling of the $5d$ $Ir^{4+}$ ion, lends itself to a wide array of potential electronic and magnetic states. In this paper, we present longitudinal field and zero field µSR measurements of $Er_2Ir_2O_7$. The µSR response of $Er_2Ir_2O_7$ is dominated by the strong fluctuating fields of the $Er^{3+}$ ($J = 15/2$) moments, showing a peak in fitted relaxation rate, $\lambda_2$, at $T = 15$ K . The µSR spectra show just weak signatures of the Ir transition, that were easily suppressed in applied longitudinal field, however full decoupling of the muons from local magnetic fields was not achieved even in applied fields up to 3 kG.

## *Introduction*

For more than a decade, the pyrochlore oxides, of a general formula $A_2T_2O_7$, have been attracting considerable attention within the condensed matter community for their frequently exotic and non-trivial physical properties.[1] The respective ionic radii of typically rare-earth ($A^{3+}$) and $d$-metal ($T^{4+}$) elements play a crucial role in the stability of the crystal structure (space group $F d$ -$3 m$), eventually leading to lower symmetry structures or/and disorder of the elements in the lattice.[2] The character of magnetic moments of respective ions, being localized or itinerant, together with the geometrically frustrated lattice, govern the magnetic properties/ground states of the oxides. The all-in-all-out (AIAO) long-range order,[3,4] partially ordered systems,[5] quantum order-by-disorder transition,[6] fragmentation of magnetic moments,[7] unconventional spin glass,[8] spin ice,[9] ordered spin ice,[10] chiral spin liquid state,[11] or structural-disorder-induced quantum spin liquid,[12] represent just some of the exotic properties reported for $A_2T_2O_7$. Moreover, a delicate balance between exchange, dipolar and spin-orbit interactions provides a ground for topologically non-trivial phases such as a Weyl semimetal state, Fermi-arc surface states, or topological band insulators.[13–16]

Furthermore, both $A$ and $T$ sublattices being concurrently magnetic may also have a strong coupling between them, leading, in turn, to more complex magnetic properties. Focusing on the pyrochlore iridates, the subject of the present paper, both sublattices, $A$ and Ir, (can) order magnetically, influencing each other.[3,17] The magnetocrystalline anisotropy, induced by Ir ordering, and $f$-$d$ electron-electron exchange interaction in $A_2Ir_2O_7$ strongly influence the magnetic order of the $A$ sublattice (e.g. $Er_2Ir_2O_7$ and $Tb_2Ir_2O_7$,[18,19]) and are put forward as the cause of giant magnetoresistance in $Nd_2Ir_2O_7$.[20] The nature of the iridium sublattice ordering, with ordering temperature increasing from 120 K to 150 K almost linearly with $A^{3+}$ ionic radius from $A$ = Sm to Lu, has previously been discussed.[21,22] The majority of studies suggest the AIAO order, while a few ascribe the high-temperature properties to short-range correlations between $Ir^{4+}$ magnetic moments.[19,20,23] In this aspect, we highlight two recent powder-neutron-diffraction studies: Guo et al.[3] determined the AIAO order of both Ir and Tb sublattices in $Tb_2Ir_2O_7$ below 125 K, where the Ir sublattice (moment of 0.55 $\mu_B$) induces the order of Tb sublattice. The second (XY) component of the Tb magnetic moments, with substantially larger value of the moment (5.22 $\mu_B$), although still reduced compared to free ion value, was followed at lower temperature (approximately 10 K). Jacobsen et al.[24] observed two magnetic peaks in difference diffraction patterns, measured below and above Ir ordering temperature, of $Yb_2Ir_2O_7$, and more importantly $Lu_2Ir_2O_7$. The peaks were identified to be consistent with the AIAO magnetic structure with Ir$^{4+}$ magnetic moments of 0.45 $\mu_B$. Below 2 K, a ferromagnetic ordering of Yb$^{3+}$ moments was observed. A reduced value of magnetic moment, reaching 0.9 $\mu_B$ at 40 mK, was interpreted to be a consequence of a magnetic phases' competition caused by a coupling to the iridium AIAO ordered sublattice. Somewhat surprisingly, the neutron diffraction studies on Ho$_2$Ir$_2$O$_7$ and Er$_2$Ir$_2$O$_7$ did not reveal any magnetic signal related to the Ir sublattice.[7,19] Nevertheless, the AIAO magnetic structure of the Ir$^{4+}$ and $A^{3+}$ magnetic ions sitting in the corners of the respective tetrahedrons (subjected to a strong [111] anisotropy) was predicted by the symmetry analysis.[25] Therefore, considering the other $A_2$Ir$_2$O$_7$ analogues, the same AIAO ordering of the Ir sublattice, reported on in Lu$_2$Ir$_2$O$_7$ and Yb$_2$Ir$_2$O$_7$, is anticipated also in those pyrochlores.

The character of magnetic ordering of the iridium sublattice, as well as the rare-earth sublattice, being long-range or short-range, has frequently been determined/estimated based on a bifurcation of zero-field-cooled (ZFC) and field-cooled (FC) dc-magnetization,[19,21,26,27] and the presence/absence of a magnetic signal in neutron diffraction data,[3,7,19,24] as listed above. Dynamical studies have been performed mainly on the light-$A$ counterparts, nevertheless leading to important results: No magnetic transition connected to the Ir sublattice was reported for Pr$_2$Ir$_2$O$_7$, while the Pr sublattice reveals exotic magnetic behaviour possibly explained by either nanosecond-scale fluctuations of magnetic moments in a long-range ordered state, or crystal field level splitting induced by a $\mu^+$-lattice distortion.[28] The long-range magnetic ordering of both Ir (below 33 K and 117 K) and $A$ sublattices (below 10 K) was deduced analysing µSR data of Nd$_2$Ir$_2$O$_7$ and Sm$_2$Ir$_2$O$_7$.[29] The magnetic moments on the two sublattices, having magnitudes of the order of 0.1 $\mu_B$, couple ferromagnetically in the Nd analogue and antiferromagnetically in the Sm analogue. The AIAO ordering was confirmed also by DFT and dipole-field calculations to model the local fields observed by implanted muons for each likely magnetic ground-state. Eu$^{3+}$ with $J$ = 0 makes Eu$_2$Ir$_2$O$_7$ an ideal system for a study of Ir$^{4+}$ magnetism. Long-range commensurate magnetic order, however also with anomalously slow spin fluctuations, were reported for this iridate.[30] A short-range ordering of Dy moments was speculated about in Dy$_2$Ir$_2$O$_7$.[31] In contrast, long-range order of Ir moments was concluded investigating Yb$_2$Ir$_2$O$_7$,[32] which was recently confirmed also by neutron diffraction experiment.[24] The same AIAO order was reported also for Lu$_2$Ir$_2$O$_7$, while the dynamical response indicates a rather short-range order.[23,24] Interestingly, characterisation of Y$_2$Ir$_2$O$_7$ suggested that the bifurcation of ZFC and FC magnetization measurements, the metal-insulator transition, and Ir long-range ordering are not necessarily occurring at the same temperature,[32] as reported for other analogues.

Deeper understanding of the magnetic ordering of the Ir sublattice, its systematics, development with $A$ substitution, and its effects (molecular field, $f$-$d$ exchange, spin-orbit coupling) on the rare-earth sublattice is, leastwise, incomplete. Our present muon-spin-resonance (µSR) experiments performed on Er$_2$Ir$_2$O$_7$, confirm the scenario of mostly dynamic spins, showing large muon depolarisation at all temperatures and minimal asymmetry recovery with application of a 3 kG longitudinal field.

## Experimental details

The investigated $A_2$Ir$_2$O$_7$ pyrochlores were synthesized from a stoichiometric mixture of initial $A_2$O$_3$ and IrO$_2$ oxides (99.99% pure, metals basis, AlfaAesar) employing the CsCl (99.999% pure) flux method. The mixture was repeatedly reacted for 12 hours at 1073 K in air, with a regrind after each reaction cycle. Further details on samples' preparation are reported elsewhere.[23,33,34] The prepared sample was washed of CsCl and characterized by means of powder x-ray diffraction and electron microscopy. The final samples contained, beside the majority stoichiometric pyrochlore $A_2$Ir$_2$O$_7$ phase, also a contribution (up to 3%) of unreacted $A_2$O$_3$ and elemental Ir.[23,33,34] The presence of secondary phase/s is common whilst preparing iridate pyrochlore oxides, as evaporative losses of volatile iridium oxides is difficult to accurately account for, however their impact on physical properties is considered only minor.[3,11,19,24,35] The DC magnetization and specific heat measurements, and in the case of $A$ = Er neutron scattering experiments, were recently performed on samples investigated in the present paper,[22,23,33,36] documenting their good quality, as well as revealing complementary properties to our present µSR results.

Zero-field, ZF, and longitudinal-field, LF, (of up to 3 kG) muon spin spectroscopy (µSR) experiments were performed at the Rutherford Appleton Laboratory (RAL, ISIS), Didcot, using the MUSR spectrometer.[40] The ~2.5g sample of polycrystalline Er$_2$Ir$_2$O$_7$ was packed in a standard µSR Ag plate sample holder with a small amount of dilute GE-varnish for improved thermalisation, and covered by an Ag foil. Measurements in the range 2.5 K < $T$ < 240.0 K were performed in a Variox, He exchange cryostat, whilst a $^3$He sorption cryostat was used for measurements in the range 0.3 K < $T$ < 20.0 K. The asymmetry was corrected for detector efficiencies and sample-environment attenuation using a high-temperature TF20 (transverse field, 20 G) measurement, separately for each cryostat setup.

## Experimental results

The magnetism of both Er$^{3+}$ and Ir$^{4+}$ in Er$_2$Ir$_2$O$_7$ compound contributes to the dynamic response of the system, each of them in possibly different time scales with mutual influence on each other. Muons implanted into the sample are sensitive to the internal spin dynamics at the implantation site(s). This local probe is especially powerful as the internal response can be probed even in zero applied external field, and is sensitive to dynamics in the MHz regime. Oscillations in the time response of muon asymmetry are typically observed in an ordered magnetic state, as is seen in $A_2Ir_2O_7$ ($A$ = Yb, Eu, Nd, Y).[32,41,42] In the case of the iridates where two magnetic sublattices are present, and if two muon implantations sites are observed, the application of an applied longitudinal field holds the possibility to effectively decouple muons experiencing a weak local field, enhancing sensitivity to the second, stronger local field, implantation site.

At all temperatures and fields there is a strong depolarisation of the implanted muon ensemble. A single function was used to parameterise the time-dependent depolarisation of the initial muon asymmetry, $A(t)$, at all temperatures and fields, incorporating a combination of two Lorentzian relaxation functions, as well as a time-invariant background contribution accounting for muons stopping within the silver sample holder, resulting in a total fit function of

$$A(t) = A_1 e^{(-\lambda_1 t)} + A_2 e^{(-\lambda_2 t)} + A_b, \quad (1)$$

where $A_1$ and $A_2$ represent the magnitudes of the relaxing components, $\lambda_1$ and $\lambda_2$ are the depolarisation rates of each, and $A_b$ is the time-independent background. Selected data and fits, are shown in Fig. 1. For all fits, the background term was fixed through a temperature series ($A_B$ = 5.6 - 5.8 %, depending on cryostat and applied field), determined by an average of linear fits of the high-time region, 10 μs < $t$ < 32 μs, where the asymmetry is constant. The large fluctuating magnetic moment of the $Er^{3+}$ ion leads to a rapid depolarisation of the muon spin, as the fluctuating dynamics are within the muon experimental time scale. An increase in the relaxation rate of the fast relaxing component, $\lambda_1$, is observed on cooling below ~115 K (60 K) for zero-field (3 kG longitudinal field) measurements (Fig. 2 (b)) as the continued slowing of the $Er^{3+}$ moment dynamics causes the muon spin relaxation to move outside of the time window of the experiment (~10 MHz). To compensate for this, the remaining tail of this fast-relaxing component was accounted for by fixing the component's relaxation rate to 10 MHz below these temperatures, allowing a good fit of the slower relaxing component, $\lambda_2$.

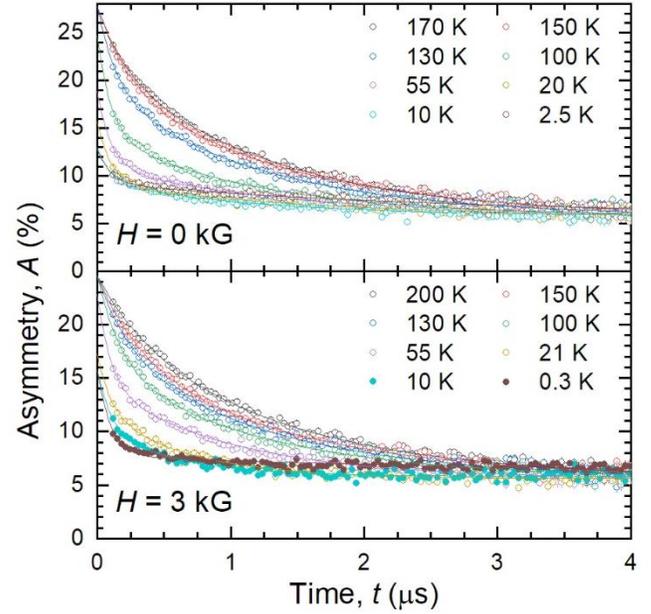

FIG. 1. μSR spectra at various temperatures; solid-line fits to the two-component function described in the text (1) in (a) ZF, and (b) with a 3 kG LF.

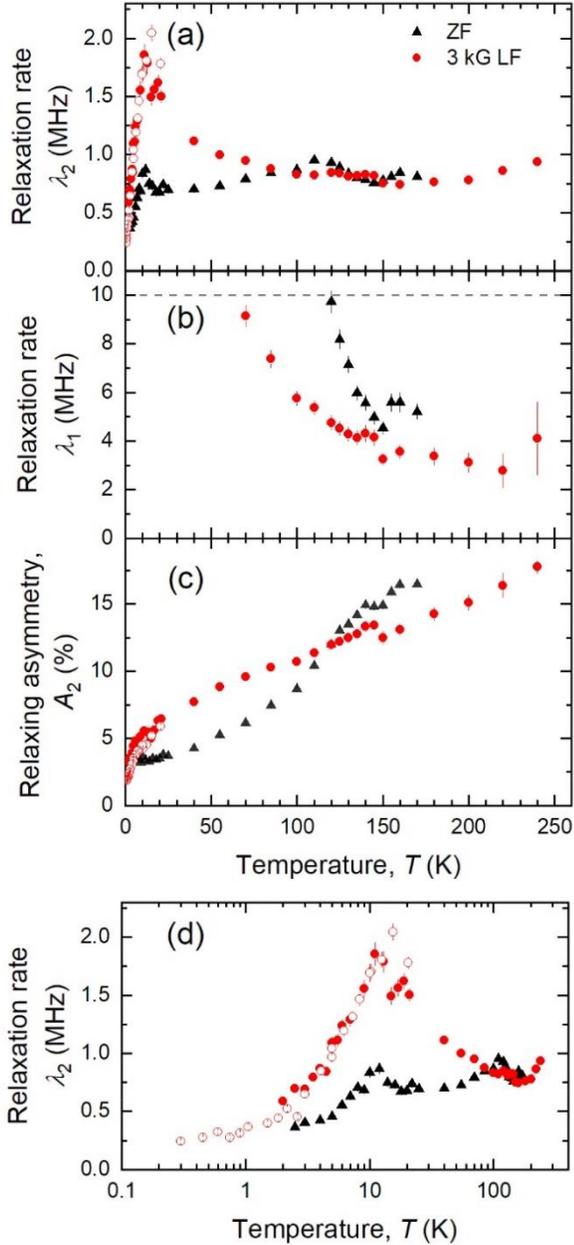

Figure 2. Fitted parameters from the μSR spectra of $Er_2Ir_2O_7$. Open symbols represent the sorption fridge measurements, closed symbols the measurements using the He-flow cryostat. Panels (a) and (b) show the relaxation rate of the slow, and fast relaxing components $\lambda_2$ and $\lambda_1$, respectively. The dashed line in (b) represents the maximum relaxation rate reasonably observed within the pulsed-source spectrometer. Panel (c) displays the fitted relaxing asymmetry of the slow relaxing component, $A_2$. Panel (d) depicts the relaxation rate, $\lambda_2$, against a logarithmic temperature scale to highlight the low temperature behaviour.

Considering the zero-field data, an anomaly can immediately be seen in the magnitude of the slow-relaxing component, $A_2$ (Fig. 2 (c)), at 140 K, associated with the onset of the metal-insulator transition and a magnetic transition of the Ir sublattice. Unlike several other members of the $A_2Ir_2O_7$ family, where oscillations in the μSR spectra immediately signify static long-range order,[3,32,42] the response of $Er_2Ir_2O_7$ appears more glassy in nature. No oscillations are observed, and the slow-relaxing component, $\lambda_2$, shows a weak broad peak with onset at 140 K but centred at ~115 K (Fig. 2 (a)). On cooling below this peak, the relaxation rate decreases slowly from $\lambda_2 = 0.952(16)$ MHz at 110 K to 0.67(3) MHz at 18 K, followed by an increase to a second peak centred at 15 K (Fig. 2 (d)). This 15 K peak in $\lambda_2$ can, at first glance, be attributed to a glassy magnetic transition, with Er spin-dynamics slowing into and then continuing to slow out of the μSR time-window. However, this peak at 15 K is close in temperature to an anomaly seen in specific heat data, previously attributed to $Er^{3+}$ crystal field excitations, at ~25 K, consistent with the crystal field splitting scheme determined through inelastic neutron scattering.[33] Below 15 K, the relaxation rate of this slow-relaxing component drops steadily with no additional anomalies down to the base temperature of the zero-field experiment, 1.5 K.

In order to estimate the internal field experienced by the muon, $H_{int}$, longitudinal field dependent μSR spectra were also collected at a fixed temperature of 5 K. Typically, as the applied longitudinal field increases, it decouples the muon from the internal field, resulting in an increase in the muon initial and baseline asymmetry with a step-like shape centred at the internal field value.[43,44]

Such behaviour has been observed in other $A_2Ir_2O_7$ spectra, resulting in a fitted internal field of $H_{int}$ ~500 G, 950 G and 1130 G, for $A$ = Nd, Y and Yb, respectively.[32,41] For $Er_2Ir_2O_7$, however, as is seen in Fig. 3, the fitted relaxing asymmetry, $A_2$, increases sharply above $H_{LF}$ = 1000 G but is not saturated even up to the maximum field of the spectrometer, $H_{LF}$ = 3000 G, preventing fitting but suggesting an $H_{int}$ > 2000 G at 5 K.

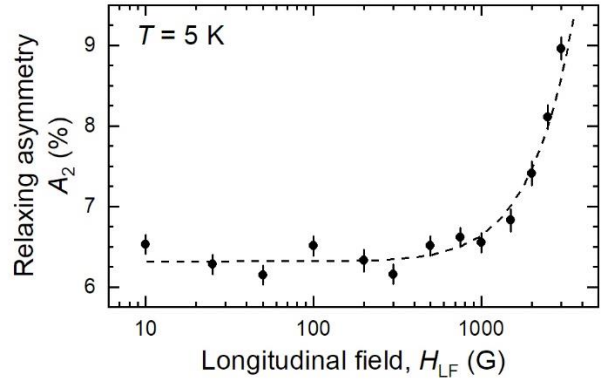

FIG. 3. Fitted relaxing asymmetry, $A_2$, as a function of applied longitudinal field at 5 K. Dashed line is a guide to the eye, highlighting the onset of decoupling from the internal field.

Fits of the temperature dependent spectra in an applied 3 kG longitudinal field (Fig. 1 (b)) shares several features with that of the zero-field data. Again, the relaxation rate of a fast relaxing component, $\lambda_1$, increases rapidly with decreasing temperature, and exceeds 10 MHz below 60 K. Unlike the zero-field data, the signature of the Ir MIT at ~140 K is significantly supressed in the response of the slow relaxing component, $\lambda_2$, although a step-like feature is still observed in the fitted relaxing asymmetry, $A_2$, at 145 K. Unlike the high temperature anomaly, the peak in $\lambda_2$ centred at ~15 K is more pronounced in applied field. Suppression of the high temperature peak suggests the field effectively decouples muons sensitive to the iridium local fields, whilst enhanced relaxation rates of the 15 K feature imply it is primarily associated with dynamics of the $Er^{3+}$ moment.

## *Discussion*

Although there is no observable oscillating component in the µSR response, as was previously observed in $A_2Ir_2O_7$ ($A$ = Yb, Eu, Nd, Y),[32,42,50] it remains possible that an oscillation from Ir-sublattice ordering is present but highly damped by the large fluctuating Er fields such that it is no longer observed by the start of the pulsed source spectrometer time window ($t$ > 0.2 µs). The zero field response of $Er_2Ir_2O_7$ at high temperatures shares similar features to that observed in $Yb_2Ir_2O_7$,[32] with a small peak in the relaxation rate of the slow relaxing component close to the MIT, indicating a change in the spin dynamics. Such a feature is observed when $A^{3+}$ is also non-magnetic ($A$ = Y), confirming it origins in coupling to spin dynamics of the Ir sublattice.[32]

Due to the significantly larger moment of $Er^{3+}$ ($J$ = 15/2) compared to $Ir^{4+}$ ($J_{eff}$ = 1/2), the local field that the muon experiences from the two ions is expected to be significantly different. Indeed, the application of a 3 kG longitudinal field was enough to fully decouple the muons from the $Ir^{4+}$ local fields, supressing sensitivity to the change in spin dynamics at the high temperature transition. Decoupling from the $Ir^{4+}$ local field increases sensitivity to the continuing changes in $Er^{3+}$ spin dynamics, with an enhanced peak on cooling below 15 K.

With regards to the observation of two muon stopping sites – previous DFT calculations of electrostatic potentials in $A_2Ir_2O_7$ iridates suggest a singular crystallographic stopping site, although studies are not in agreement on the specific location.[51–53] Indeed, spectra of a closely related hafnate were also best modelled by a two-lorentzian component fit.[54] Given the preference for muons to implant close to the negatively charged oxygen anion in oxide materials (typically $r$~1 Å from $O^{2-}$ species [55]), and the presence of two crystallographically distinct oxygen sites in the pyrochlore structure, a two-component scenario involving different local fields is not immediately surprising. The relative occupation of these sites is not expected to significantly change across the temperature range investigated here, yet we find a continuous shift of spectral weight from the slow-relaxing component, $A_2$, to the fast relaxing component, $A_1$, at least with the time resolution of the used spectrometer. This observation leads us to believe that the two components do not necessarily represent different crystallographic sites, but instead an inhomogeneous local field distribution that is simply modelled best by two distinct components. Such an inhomogeneous field distribution scenario is consistent with a complex magnetic transition at the Ir MIT, involving static, dynamic and short range components, and has previously been seen in other iridium systems with multiple magnetic lattices.[56,57]

On cooling from high temperatures, first there is transition within the $Ir^{4+}$ sublattice in the temperature range from 120 K to 150 K, linearly dependent on $A^{3+}$ ionic radius.[21,22] Bifurcation of zero-field-cooled and field-cooled dc-magnetisation data confirms at least some component of static magnetism,[19,21,26,27] which is confirmed as *long-range* order in $A_2Ir_2O_7$ ($A$ = Yb, Eu, Nd,Y) by oscillations in µSR spectra, as well as additional neutron diffraction peaks in $A$ = Lu, Yb, Tb consistent with an AIAO structure.[3,24] However, simple static long range order does not account for the observed continued dynamics of our and previous µSR spectra. As such the transition must contain a significant dynamical component. Muons, sensitive to fluctuation times in the $10^{-6} – 10^{-12}$ s range, may observe a slow-fluctuation ($\tau_0$ > $10^{-6}$ s) spin glass as static order, resulting in the observed oscillations in the muon spectra. Variation in the local field throughout a sample would give rise to the apparent inhomogeneous characteristics observed in $Er_2Ir_2O_7$.

On further cooling, a coupling of the muon to $Er^{3+}$ crystal field excitations is observed in the µSR response of $Er_2Ir_2O_7$, evidenced by a peak in muon relaxation rate at 15 K. This temperature is in line with a feature previously observed in analysis of specific heat data for $Er_2Ir_2O_7$,[33] and is consistent with evidence of correlated magnetic behaviour of the $Er^{3+}$ sublattice in isothermal magnetisation measurements below 20 K.[33] Additional dynamical measurements of this intriguing pyrochlore iridate family using muons from pulsed-sources, with higher time resolution, are needed to further clarify the nature of the complex competing static and dynamic behavior observed.

## *Conclusions*

In summary we have measured zero-field and longitudinal field (3 kG) µSR spectra for $Er_2Ir_2O_7$. Fits to the µSR spectra show weak evidence of the Ir magnetic transition at ~140 K, although no oscillations that would confirm long range magnetic order were observed in the pulsed-source data. The strongly relaxing muon response was, however, most sensitive to the strong fluctuating field of the $Er^{3+}$ ($J = 15/2$) moments, showing a peak indicative crystal field excitations at $T = 15$ K. The muon relaxation could not be fully decoupled from the sample dynamics, even in the maximum applied field (3 kG), however the signatures of the Ir magnetic transition were heavily suppressed in our longitudinal-field temperature-dependent measurements. Coupled with the previous observation of (at least partial) long-range order in many $A_2Ir_2O_7$ oxides, these measurements highlight the unconventional nature of the electronic state in the rare-earth pyrochlore iridates, certainly warranting further study.

## *Acknowledgements*

We would like thank Dr Adrian Hillier and Dr Peter Baker for useful additional advice and discussion during the MUSR experiment, as well as the UK Science and Technology Facilities Council (STFC) for access to muon beamtime at ISIS, MUSR facility.[40] The preparation, characterization and measurement of bulk physical properties on $A_2Ir_2O_7$ samples were performed in MGML (http://mgml.eu/), which was supported within the program of Czech Research Infrastructures (project no. LM2018096).

Additionally, this work was supported by the Czech Science Foundation under Grant No. 18-09375Y. The work of K.V. was further supported by GAUK Project Number 558218.